\documentclass[%
 preprint,
 amsmath,
 amssymb,
 aps,
]{revtex4-1}

\usepackage{cmbright}
\DeclareFontShape{OT1}{cmss}{m}{it}{<->ssub*cmss/m/sl}{}

\usepackage[font={small,stretch=1.05}]{caption}
\usepackage[document]{ragged2e}
\usepackage[none]{hyphenat}
\usepackage [english]{babel}

\usepackage{graphicx}
\usepackage{dcolumn}
\usepackage{bm}
\usepackage{gensymb} 
\usepackage{caption}
\usepackage{subcaption}
\usepackage{stmaryrd}
\usepackage{amsmath}
\usepackage[utf8]{inputenc}
\usepackage{hyperref}

\usepackage[margin=1in]{geometry}
\usepackage{chngcntr}
\counterwithin{table}{section}
\usepackage{tabularx}
\usepackage[T1]{fontenc}
\usepackage{etoolbox}
\AtBeginEnvironment{tabular}{\fontfamily{phv}\fontsize{7}{8}\selectfont}{}{}

\usepackage{titlesec}
\titleformat{\section}[block]{\normalfont\Large\bfseries}{\thesection}{1em}{}
\titleformat{\subsection}[block]{\normalfont\large\bfseries}{\thesubsection}{1em}{}
\titleformat{\subsubsection}[block]{\normalfont\normalsize\bfseries}{\thesubsubsection}{1em}{}

\begin{document}

\title{Beaded metamaterials}

\author{Lauren Dreier$^{1}$, Trevor J. Jones$^{2}$, Abigail Plummer$^{3}$, Andrej Košmrlj$^{3,4}$, P.-T. Brun$^{2}$ }

\affiliation{$^1$School of Architecture, $^2$Department of Chemical and Biological Engineering, $^3$Department of Mechanical and Aerospace Engineering, $^4$Princeton Materials Institute, Princeton University, Princeton, New Jersey 08540, USA
}

\begin{abstract} 
From the pragmatic to the symbolic, textiles play a prominent role in some of the most demanding yet ubiquitous scenarios, such as covering the complex and dynamic geometries of the human body.
Textiles are made by repeated manipulations of slender fibers into structures with emergent properties~\cite{hearle2008physical,banerjee2014principles}. Today, these ancient metamaterials are being examined in a new light\cite{sanchez20233d,singal2024programming}, propelled by the idea that their geometric structures can be leveraged to engineer functional soft materials\cite{bertoldi2017flexible,reis2015buckle,xiong2021functional,sanchez2021textile}. However, per their inherent softness, textiles and other compliant materials cannot typically withstand compressive forces~\cite{wang2021structured}. 
This limitation hinders the transfer of soft matter's rich shape-morphing capabilities~\cite{xia2022responsive} to broader research areas that require load-bearing capabilities. Here we introduce \textit{beading} as a versatile platform that links centuries of human ingenuity encoded in the world of textiles with the current demand for ‘smart’, programmable materials.
By incorporating discrete rigid units, i.e. \textit{beads},  into various fiber-based assemblies, beadwork adds tunable stiffness to otherwise flaccid fabrics, creating new opportunities for textiles to become load-bearing. We select a shell-like bead design as a model experimental system and thoroughly describe how its mechanics are captured by friction, the material properties of the constituent elements, and geometry. The fundamental characterization in this study demonstrates the range of complex behaviors possible with this class of material, inspiring the application of soft matter principles to fields that ultimately demand rigidity, such as robotics and architecture.
\end{abstract}

\maketitle
\section{Introduction}

Worn close to the body or stretched overhead as a roof, textiles are known for their softness and versatility. Their tendency towards compliance can be explained by the fibers or yarns they are made from, which bend easily due to their slender geometry. Geometry also contributes to textile mechanics when fibers are manipulated — by hand, machine, or nature — into hierarchical structures~\cite{banerjee2014principles}, giving way to new and sometimes surprising properties. For example, inelastic yarns can become effectively elastic by knitting them into a stretchable fabric, and discrete layers of wool fabric can cohere without glue or stitching by felting. A material that exhibits this kind of emergent, global behavior by leveraging its structure is known as a \textit{mechanical metamaterial} \cite{bertoldi2017flexible}. Classified as such, materials as ancient and commonplace as weaves and knits are currently being examined in a new light, propelled by the recent paradigm shift towards understanding and ultimately harnessing nonlinear mechanics such as elastic buckling \cite{reis2015buckle,holmes2019elastic} and plasticity \cite{keim2019memory}. 
Progress has been made in characterizing the relationship between mechanical properties and fiber topology, including knits \cite{poincloux2018knit}, knots \cite{patil2020knots,Jawed2015knots}, yarns\cite{warren2018clothes}, woven lattices\cite{baek2018gridshells,poincloux2023indentation,yu2021numerical}, and non-woven fiber networks\cite{kabla2007fiber,andradesilva2021nest,dumont2018emergent}, in some cases using actuation (e.g., stimuli-responsive matter, pneumatics) to create functional shape-morphing matter with promising applications in wearable electronics\cite{xiong2021functional,sanchez2021textile} and soft micro robotics\cite{chen2019controlled}. Traditional handicrafts with 2D hierarchical geometry have also been reframed as mechanical metamaterials, including origami, kirigami\cite{zhai2021mechanical}, and pleating-like corrugations\cite{meeussen2023multistable}. Like rod- and fiber-based materials, these systems also exhibit emergent mechanical properties and show potential for shape programming.

A major challenge in employing shape-morphable materials in real-world engineering scenarios lies in eliciting rigidity from materials that are otherwise soft, i.e. flexible enough to change shape without damage \cite{kuder2013variable,blanc2017flexible,hao2022low}. 
Granular systems close to the jamming transition demonstrate this dual behavior \cite{liu1998jamming,majmudar2007jamming}. 
Jamming can be paired with compliant structures such as flexible membranes \cite{bakarich2022pump, brown2010universal, wang2021structured} and fiber elements \cite{guerra2021emergence} to gain functionality. 
However, jamming elements is typically a stochastic process and requires a positive form to prescribe shape.  
Tensegrity structures provide another framework for accomplishing rigid shape morphing using struts and cables, though systematic design remains challenging \cite{liu2019tensegrity, shah2022tensegrity}.
To achieve tunable rigidity in a programmable, ordered platform, in this study we take up \textit{beadwork}, an ancient textile system practiced by cultures around the world that links discrete rigid elements, i.e. beads, via a compliant thread that prescribes topological order. Beadwork combines rigid, volumetric bead elements that provide resistance to compressive loads with cohesive force provided by the thread, which allows beaded objects to take a fully programmed shape without formwork or an enclosing membrane.

\begin{figure}[!h]
    \includegraphics[width=\textwidth]{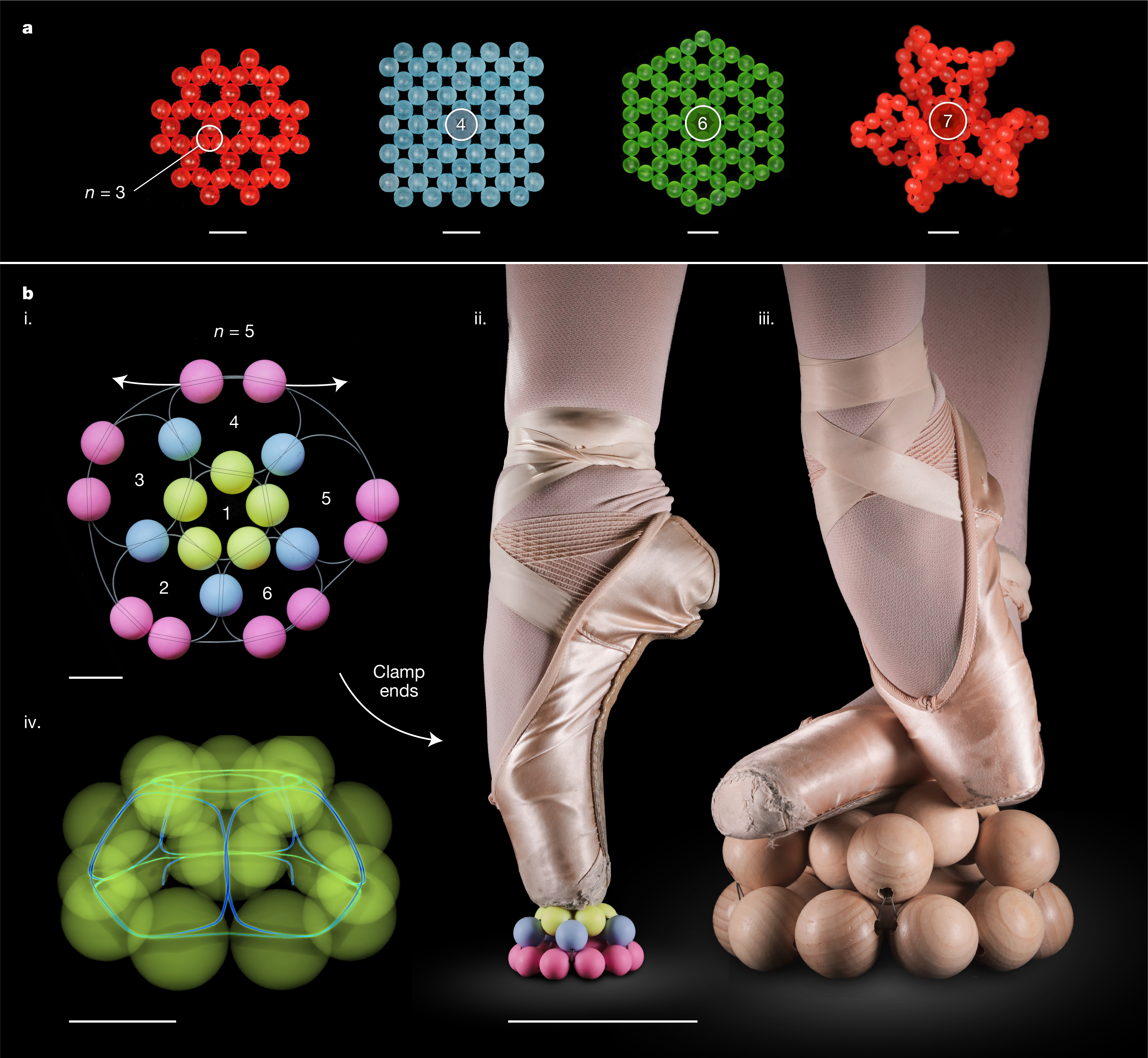}
    \caption{
    \textbf{Design and emergent stiffness of objects beaded using an angle weave technique.}
    \textbf{a}, Sample angle-woven swatches comprised of adjoining loops of $n$ beads, where $n = 3$--$7$. Non-Euclidean tiling of unit cells can lead to three-dimensional surfaces. Scale bars, 20~mm. 
    \textbf{b}, A beaded shell shown before tensioning (i), with thread pattern schematically traced over a photograph of otherwise opaque, 17-mm acrylic beads threaded with steel rope. Scale bar, 20~mm. Applying tension to the two free ends sends the half-dodecahedral shell out-of-plane (ii--iv). The structure becomes rigid across length scales by clamping the thread, capable of bearing large ``pointe'' (ii) and distributed (iii) loads. Scale bar, 100~mm.
    (iv), $\mu$-CT reconstruction of the model of the same design using 10~mm acrylic beads (yellow) threaded with 0.25-mm nitinol wire (blue). Scale bar, 10~mm.
    }
    \label{fig:fig1}
\end{figure}

Spanning simple 1D ‘string of pearls’ configurations to fully 3D interlaced weaves, beadwork has no single topological form but draws on many fiber assembly techniques to create a wide range of designs. In this work, we study \textit{angle weave} stitches due to their versatility in creating 2D and 3D designs (see Fig. \ref{fig:fig1}a)\cite{chuang2012molecular,fisher2012tiling,fisher2015triangles,tsoo2017poly}. In angle weaves, beads are strung in adjoining loops, forming rings of $n$ beads that can cover a plane or 3D surface. Applying tension to the bead thread minimizes excess lengths between bead elements. The beaded object thus takes a shape prescribed by the discrete tiling pattern of angle weave building blocks. If the pattern is designed with geometric frustration, such as the heptagonal tiling shown in Fig. \ref{fig:fig1}a or the half-dodecahedral shell shown in Fig. \ref{fig:fig1}b, the beaded surface will come out-of-plane when tension is applied (see SI Video~\ref{v:1}). Here we demonstrate that these morphable textiles can be load-bearing. In Fig. \ref{fig:fig1}b we show a design for a half-dodecahedral shell comprised of loops of 5 beads. Applying tension to thread ends reduces the packing fraction and induces jamming of the granular structure. The assembly comes out of plane as the blue and yellow beads lift off the substrate. Clamping the ends in this state creates a scalable, rigid structure that can support the weight of an adult dancer -- an unusual feature for soft materials.

We devise and rationalize a series of model experiments to characterize these mechanics, using this beaded shell structure that exhibits both shape morphing and rigidity as an entry point for our study. In Fig. \ref{fig:fig1}b and SI Video \ref{v:1}, we show a 3D reconstruction of this model, obtained via micro-computed tomography ($\mu$-CT). The solid acrylic beads in yellow appear translucent, revealing the more dense nitinol wire (blue) used as thread. For a simple, 20-bead object, the image reveals a tortuous mechanical system consisting of contacts between individual beads, beads with the thread, the thread with itself, and beads with supports. Despite these complexities, experiments demonstrate a consistent and tunable mechanical response, detailed in the next section.

\section{Results}

\begin{figure}[!h]
    \includegraphics[width=\textwidth]{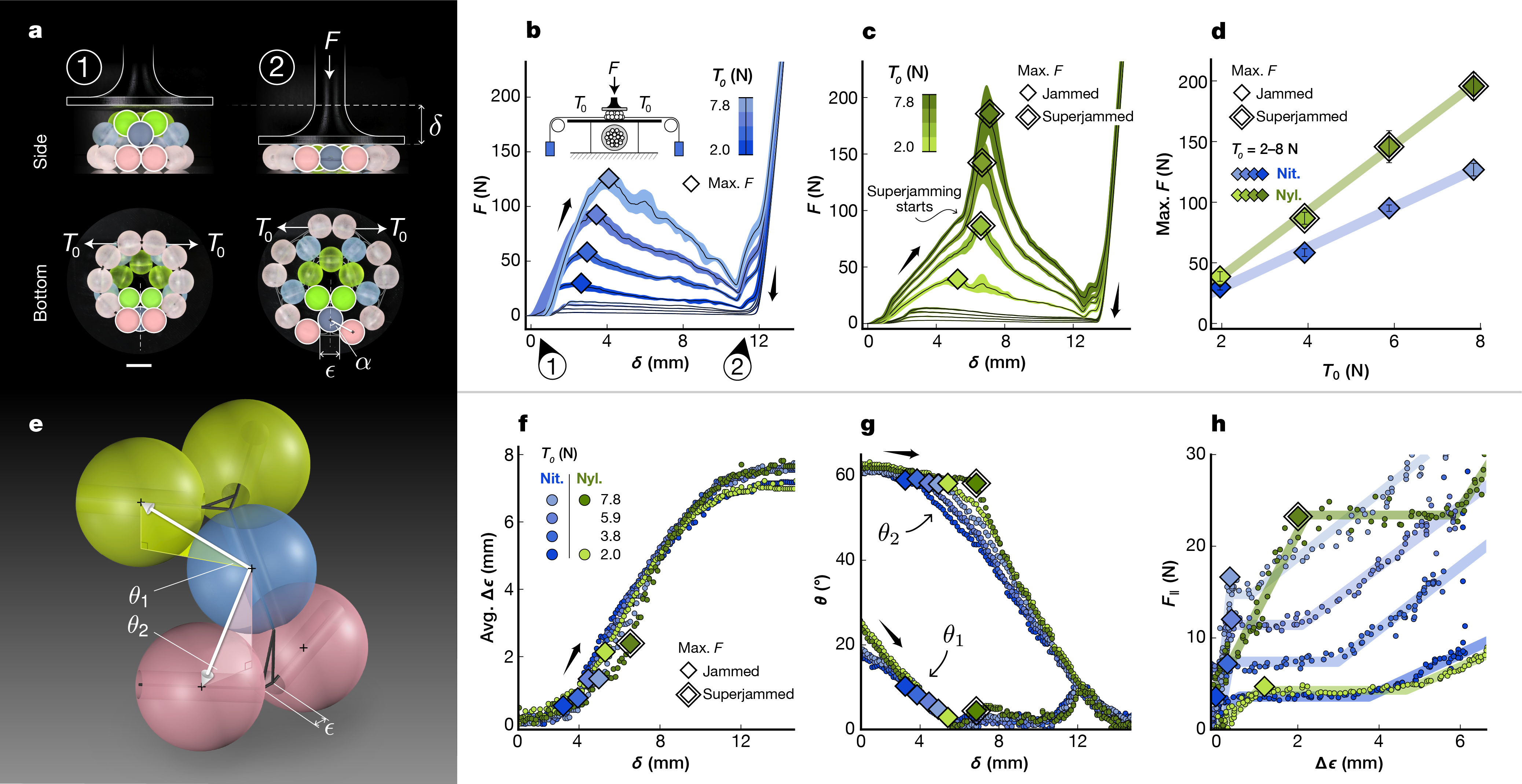}
    \caption{\textbf{A beaded shell is compressed between two plates.} \textbf{a}, Photographs of the experiment. Constant tension ($T_{0}$) is applied to thread ends as the indenter moves a distance $\delta$. Shaded regions show maximum force over all trials. Scale bar, 10~mm. \textbf{b-c}, The force response ($F$) for identical shells threaded with different materials (0.50-mm diameter nylon and 0.25-mm diameter nitinol, respectively) as a function of $\delta$, cycled three times across a range of $T_0$.
    \textbf{d}, Maximum load as a function of tension for shells beaded with nitinol (blue) and nylon (green) thread. Maximum forces that occur in a superjammed regime are outlined. 
    \textbf{e}, Illustration of a 5-bead building block that comprises the model structure. Gray lines indicate bead hole orientation. 
    \textbf{f-g}, Tracking selected beads' motion over a single cycle across a range of $T_0$. 
    \textbf{f}, In-plane dilation of the bottom layer of beads measured by the length of exposed thread ($\Delta\epsilon$) as a function of $\delta$. 
    \textbf{g}, Altitudinal angles $\theta_{1}$ and $\theta_{2}$ as a function of $\delta$. \textbf{h}, Projection of $F$ as in-plane tension between two pink beads shown as a function of $\Delta\epsilon$. A 3-part piecewise linear fit is shaded behind each trial. The return path is not shown in Figs. f-h.}
    \label{fig:fig2}
\end{figure}


In Fig.~\ref{fig:fig2} we characterize the compressive stiffness of our model system. We apply constant tension $T_0$ to the thread ends and compress the shell between two plates, as illustrated in Fig.~\ref{fig:fig2}a (see \hyperref[link:methods_shell]{Methods},  SI Video \ref{v:2}). In Figs.~\ref{fig:fig2}b-c we report the measured force $F$ vs. indenter displacement $\delta$ for identical shells threaded with different materials: (b)~nitinol (0.25~mm dia.) and (c)~nylon (1~mm dia.).
We identify several common features of the force response, as evident from the figure. First, the force to depress the shell increases until it reaches a maximum. The value of this maximum increases with the applied tension $T_0$. Second, after reaching a peak load, the shell softens so the indentation force decreases to nearly zero. The subsequent steep increase occurs when all the beads are flat on the substrate. Third, the return cycle has a large hysteresis, suggesting significant friction in the system \cite{barbier2009friction}. 
Fig.~\ref{fig:fig2}d confirms that the maximum force scales with tension. Results indicate a linear relationship between maximum force and tension across all trials. Notably, despite near-identical shells differing only in thread material, nylon trials exhibit 1.6 times greater perceived stiffness than nitinol trials, while nylon threads are comparatively softer than nitinol wires.
To understand this increased stiffness, we examine the data shown in Figs.~\ref{fig:fig2}c, where an additional, steeper slope emerges when loading nylon-threaded shells. In Extended Data Fig.~\ref{fig:fig_extdata_shells}, we study our shell across different indenter shapes, materials, and bead radii. Our results show that, in addition to thread material, this regime is sensitive to indenter shape and bead roughness. Moreover, we can facilitate conditions for this behavior regardless of the thread material by clamping thread ends (see Figs. \ref{fig:fig1}b~iii--iv, Extended Data Fig.~\ref{fig:fig_extdata_shells}d). In the clamped shells of Extended Data Fig.~\ref{fig:fig_extdata_shells}d, the slopes with nitinol and nylon
align in this regime (170 and 193~N/mm for nitinol and nylon, respectively), suggesting that mechanics shared by both shells --- such as bead-bead and bead-support contacts --- usurp thread elasticity. Accordingly, we call this regime \textit{superjamming}, owing to its physical similarity to granular jamming and accounting for the enhanced load-bearing capability of the structure. Note that shell response remains approximately the same across the material and geometric parameters we tested (See Extended Data Fig.~\ref{fig:fig_extdata_shells}a--c), aside from this superjamming regime. 

Anticipating the critical role of geometry in our problem, we quantify shape change under load in Fig.~\ref{fig:fig2}e. We identify two primary metrics: in-plane dilation $\Delta \epsilon$ of the bottom ring comprised of pink beads, and the altitudinal angles $\theta_1$ and $\theta_2$ tracking the position of yellow and blue beads. These measurements are reported in Figs.~\ref{fig:fig2}f and \ref{fig:fig2}g, respectively.
Points of maximum force are overlayed, showing the geometric conformations that correspond with peak stiffness.
Across all experiments, $\Delta\epsilon$ follows an S-curve, and the peak stiffness occurs on the lower branch. In other words, dilation of the bottom ring indicates shell softening and collapse. 
Looking at out-of-plane geometry, Fig.~\ref{fig:fig2}g shows that $\theta_1$ quickly decreases to zero regardless of the applied tension or thread material. On the other hand, $\theta_2$ remains approximately constant with increased tension. This stagnation is prolonged in the case of higher tension, causing a subtle rearrangement of the bead assembly with large consequences. In this limit, the indenter comes in direct contact with the blue beads, and the force path is vertical (see Extended Data Fig.~\ref{fig:fig_extdata_forcepath}). This choreography of deformation maximizes the force for a given displacement, thereby causing the sudden slope change noticed in Fig.~\ref{fig:fig2}c. This superjammed configuration is maintained until the structure slips.  

In Fig.~\ref{fig:fig2}h, we report the dilation force resisted by tension in the bottom ring (see \hyperref[link:methods_shell_track]{Methods}). We recover three regimes: first, the dilation force increases linearly, then plateaus near the moment the shell begins to soften, and finally increases again.
The slopes of the first regime depend on the thread material (stiffer threads yield steeper slopes) and the plateaus scale with tension. This latter observation suggests that sliding friction plays a role in the integrity of the bottom ring \cite{Persson2000Friction}. In the following section, we investigate the primary components of these observations regarding friction, geometry, and the distribution of tension throughout the network. We do this via a series of simplified experiments and reduced-order models that test and elucidate the mechanics at work in more complex beaded objects.

\begin{figure}[!h]
    \centering
    \includegraphics[width=120mm]{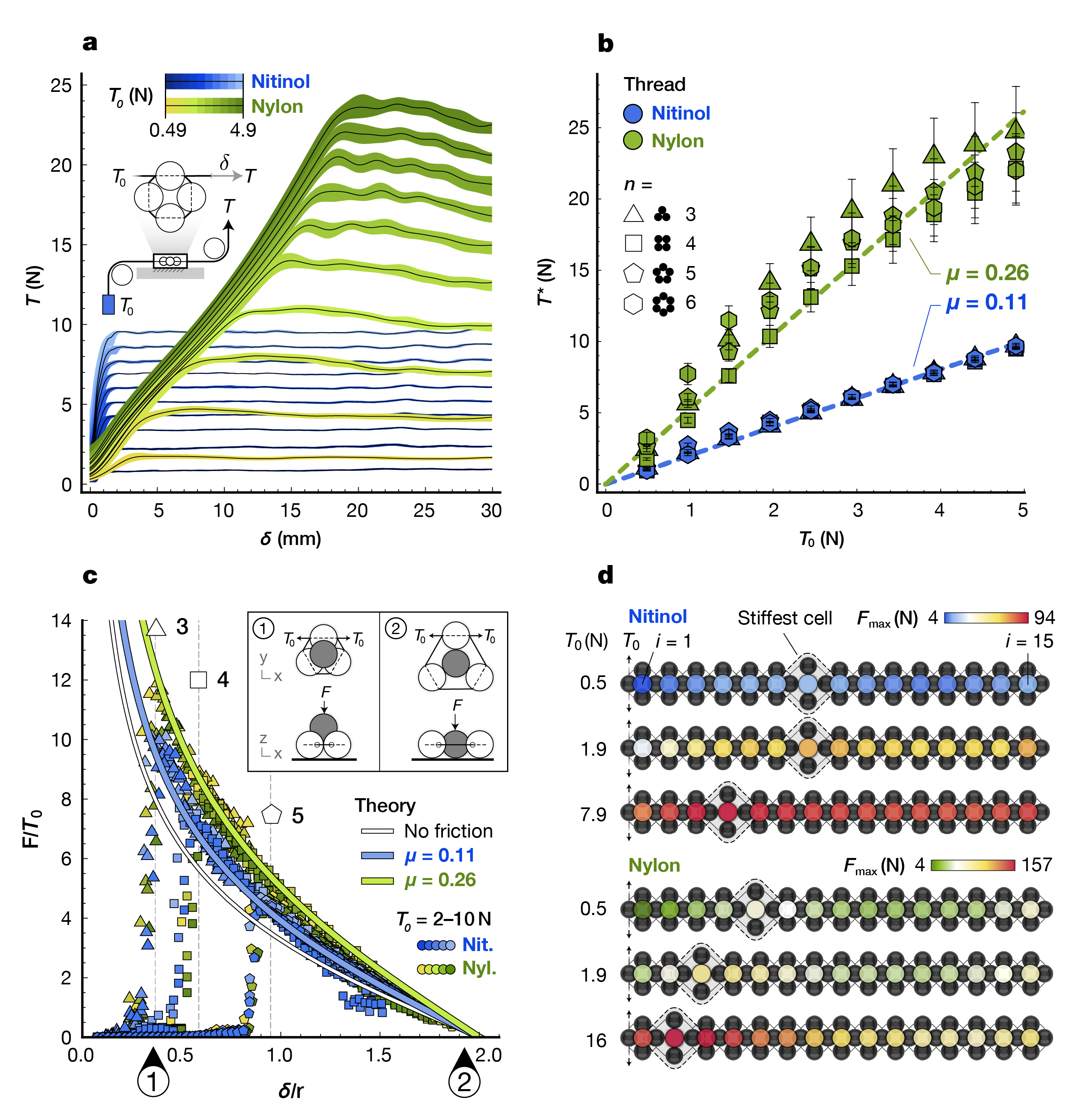}
    \caption{\textbf{Decoupling (and recoupling) friction and geometry in beaded assemblies.} 
    \textbf{a}, Measuring friction in a planar ring of $n=4$ beads. A sliding thread is pulled through a loop of constrained beads for a displacement $\delta$. Constant tension ($T_0$) is applied to one end of the thread while tension ($T$) is read at the other for a range of $T_0$ and two thread materials. 
    \textbf{b}, Kinetic friction plateaus $T^*$ plotted as a function of $T_0$ for planar bead rings where $n$ ranges from $3$ to $6$. Dashed lines indicate linear fits of the data.
    \textbf{c}, Dimensionless force ($F/T_0$) vs. indenter displacement ($\delta$) for dilating planar bead rings using a spherical-tipped indenter with the same radius as a bead. Experiments for $n=3$,$4$, and $5$ are shown as $n-$sided markers. Theory for $n=4$ and zero friction is drawn in white, with blue and green lines showing a correction for bead-thread friction in nitinol and nylon, respectively. 
    \textbf{d}, Maximum force to dilate sections of a planar chain of beads consisting of 15 adjoining rings of $n=4$ beads, threaded with nitinol and nylon for various $T_0$.}
    \label{fig:fig3}
\end{figure}


In Fig.~\ref{fig:fig3}a, we show the force required to pull a thread through a single ring of $n$ beads pretensioned by $T_0$, where $n$ has the geometric effect of imposing kinks in the thread's path (see Extended Data Fig.~\ref{fig:fig_extdata_rings}). We report the difference in tension between thread ends (see \hyperref[link:methods_dilate_exp]{Methods}). We find that the force first increases linearly and then gives way to a plateau of constant force, $T^*$. 
In Fig. \ref{fig:fig3}b, we report the value of $T^*$ that is found to scale linearly with $T_0$. The slopes are materially dependent but are found to be independent of the number of beads $n$. Leveraging this observation, we propose a first-order geometric model that ignores the thread's bending stiffness and nonlinear friction: 
\begin{equation}\label{eq:1}
T^*/T_0=e^{\mu_{\text{eff}}\phi}, 
\end{equation}
where $\mu$ is the effective frictional coefficient and $\phi=2\pi$ for all $n$ per capstan theory\cite{euler1769frottement,eytelwein1832handbuch,grandgeorge2022rod}. 
We find a fair agreement between eq.\ref{eq:1} and our data as highlighted by the collapse into lines reported in  Fig.~\ref{fig:fig3}b. We obtain $\mu_\text{eff}=0.11$ and 0.26 for nitinol and nylon, respectively. Nylon's nonlinearity at high tension is consistent with power-law friction often associated with polymeric materials, particularly in capstan configurations \cite{jung2008capstan}. 

We now allow the beads to move relative to each other and consider an experiment where a planar ring is dilated by a spherical volume (shown in gray in the inset of Fig.~\ref{fig:fig3}c). In Fig. \ref{fig:fig3}c, we report the force as we press through a ring of $n=3,4,$ and $5$ beads under constant tension. After an initial, steep increase in force when making contact (point 1 in the figure), $F$ gradually decreases to zero as the indenter aligns with the bead ring (point 2). Rescaling $F$ by $T_0$, the data collapses into a single curve per thread type. We note that $n$ does not affect $F$ beyond geometric constraints (indenter-ring contact points are shown as dashed lines, see \hyperref[link:methods_dilate_model]{Methods}).
We use geometry to predict the response by performing a force balance across the indenter and a sample bead in the ring. We assume zero friction, constant tension throughout the network, and symmetric deformation as the ring is dilated. Despite these coarse assumptions, we find fair agreement with experimental results, albeit the prediction is slightly lower than the experimental data (white line in Fig.~\ref{fig:fig3}c versus blue and green points). We correct this error by incorporating friction through Eq.~\eqref{eq:1} and the experimentally derived $\mu_{\text{eff}}$ and find excellent agreement with experiments (blue and green lines for nitinol and nylon, respectively).
Thus, we understand the deformation of a bead assembly near a free end. Next, we extend this experiment to a larger system and probe deeper into the weave. 

In Fig.~\ref{fig:fig3}d, we construct a 1D angle weave tiling that comprises 15 rings similar to those tested in Figs.~\ref{fig:fig3}a--c, but chained together with continuous thread (see \hyperref[link:methods_dilate_exp]{Methods}). We report the maximum dilation force $F_\text{max}$ for sequential dilations performed along the chain ($i=1$ denotes the ring closest to the free end). As expected, pretension $T_0$ affects the value of the maximum loads. Yet, we report that $T_0$ also dictates where peak stiffness occurs along the chain. These effects are exaggerated in nylon, where the stiffest ring approaches the free end when increasing pretension ($i=6$ and $i=2$ for $T_0 = 0.5$ or $16$~N, respectively). After the peak, $F_\text{max}$ decays and eventually plateaus at a characteristic value (approximately 15~N in the ranges we tested). In all cases, $F_\text{max}$ increases again at the boundary near $i=15$. 

We extract the dilation force $F_{\parallel}$ for each indentation performed along the chain (see Extended Data Fig.~\ref{fig:fig_extdata_chain}. Close to the free end, we recover the now familiar capstan-like plot, consisting of an initial, linear regime where the thread stretches and a plateau of constant force representing global slip, i.e., the thread slides uniformly from the free end into the structure. As one could anticipate from capstan theory, the plateau height increases with the wrapping angle and thus with $i$.  Deeper in the weave, this idealized scenario is no longer observed, suggesting that global slip does not occur. Instead, we argue that the excess length needed to open the ring is collected from the slack in the weave. As evident from Extended Data Fig.~\ref{fig:fig_extdata_rings}, the thread length surrounding each ring increases from the free end. This observation affects the mechanical behavior and, thus, the design potential of larger bead networks, which we explore in the next section.




\section{Discussion}

\begin{figure}[!h]
    \includegraphics[width=\textwidth]{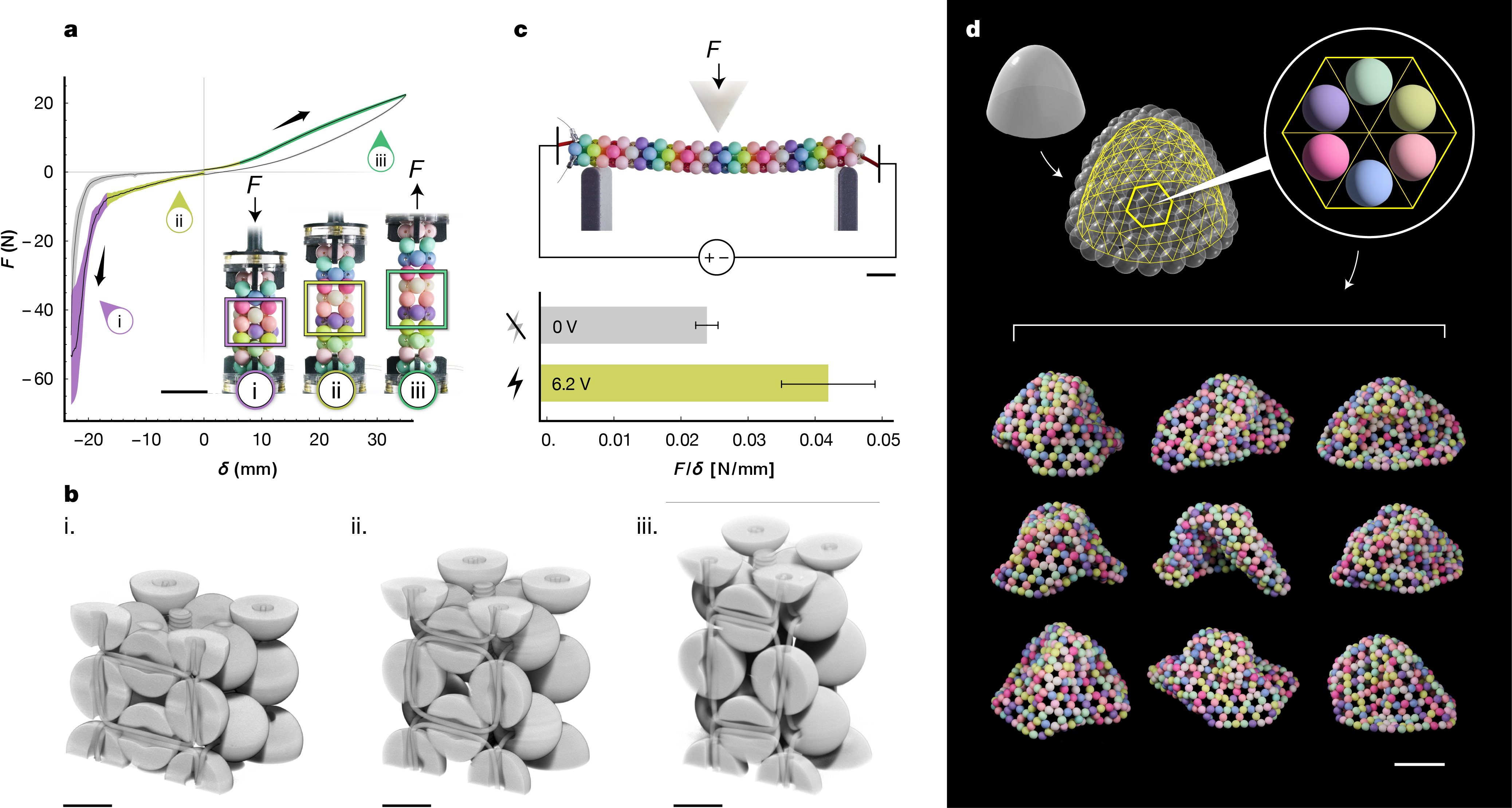}
    \caption{\textbf{Mechanical properties of larger beaded textile networks.} \textbf{a}, A beaded column consisting of loops with $n=4$ and elastomeric thread is stretched and compressed, revealing three regimes: linear elastic (yellow), strain stiffening (green), and jamming (purple). Scale bar, 30~mm.
    \textbf{b}, $\mu$-CT imaging of the column in \textit{a} with in-situ forcing in the following configurations: "jammed" (i), neutral (ii), and stretched (iii). Scale bars, 10~mm.
    \textbf{c}, A beam beaded with loops of $n=3$ is threaded with SMA wire, which contracts approx. 4\% when heated by electric current. We perform a 3-point bending test and report the mean fitted slopes ($F/\delta$) for repeated trials, with and without voltage applied. Scale bar, 15~mm.
    \textbf{d}, A bead pattern is generated by packing spheres onto an arbitrarily chosen surface (a catenary dome). The surface is constructed with large acrylic beads ($R=10$~mm), waxed polyester cord, and is subject to gravity. Bending via the distribution of characteristic slack in the network causes localized stiffening, leading to metastability and many possible rigid conformations. Scale bar, 100~mm.}
    \label{fig:fig4}
\end{figure}

In the previous sections, we have introduced beading as a material platform inspired by a tradition that brings load-bearing capability and geometric programmability to otherwise soft or disordered matter. We have focused on the mechanical response of beaded models subject to compressive loads, exemplifying that the deformation of beaded matter entails the displacement of the thread that binds discrete bead elements together. Our results demonstrate that thread elongation occurs through frictional sliding and/or stretching. We have shown that increasing tension in the thread increases contact forces between beads, leading to a stiffer elastic response upon loading. We have also shown that in the limit of large contact forces — which can be attained by leveraging geometry and material properties —  it is possible to elicit a stiff jamming regime in these materials. This regime requires the choreography introduced in Fig~\ref{fig:fig2}, where the bottom ring resists deformation, allowing the top beads to rotate into a particular jammed state. This scenario is only possible when the bottom ring, closest to the free end, is stiffer than the units above it. Fig~\ref{fig:fig4}d is consistent with this argument, showing that stiffer rings are next to the free end in the limit of large tensions. The stiffer rings are further away in the weave for smaller forces, so the bottom ring dilates first, preventing jamming. 

In Fig.~\ref{fig:fig4}a--d, we extend these concepts to other geometries and means of deformation. In Fig.~\ref{fig:fig4}a, a column beaded with elastomeric thread is stretched and compressed axially. For small displacements, we observe a linear elastic response. In compression, the beads remain in contact, so that the distance between bead centers is preserved. The thread is displaced to enable a more dense packing, where angle weave units are distorted from square-like shapes into rectangles (see Fig.~\ref{fig:fig4}b~i). Further compression yields jamming. The beads stack vertically, and recorded forces become large despite the thread’s softness. When the column is stretched, the beads move away from each other (see Fig.~\ref{fig:fig4}b~iii). After a brief linear regime where slack is used up, we observe a strain-stiffening response. Similar behaviors are typically seen in other fibrous and fiber-like materials across scales, such as knit textiles \cite{poincloux2018knit}, biological tissue \cite{winer2009elasticity}, and polymers---often described by aptly named "bead-spring" models \cite{rouse1953polymers}, based on the notion that slack exists in the system that can become exhausted; as a result, the comprising fiber-like elements are forced to stretch. In beaded structures, slack can be modified after the structure has been woven by using stimuli-responsive materials. Next, we exemplify this additional avenue of control.


In Fig.~\ref{fig:fig4}c, we bead a rod ($n=3,4$) with shape-memory nitinol wire that shrinks uniformly in length by 2--5\% when heated by electric current (see \hyperref[link:methods_larger_beam]{Methods}) \cite{dynalloy}. In a three-point bending test, we report that this relatively small strain increases the effective bending stiffness (known to scale linearly with $F/\delta$ \cite{audoly2000elasticity}) by a factor of nearly two. In this sense, coupling beading with thread that changes length provides a promising path for expanding the application space of existing fiber actuators often limited by their inherent softness \cite{liu2020filament,xu2023keratin}. 

Moreover, beading exhibits performative mechanical behavior even when working with passive materials. We find this to be particularly evident at larger scales. In Fig.~\ref{fig:fig4}d, we construct a catenary dome using 20~mm dia. beads and waxed nylon thread (see \hyperref[link:methods_larger_shell]{Methods}). In handling, we observe that bending induced by gravity triggers non-local stiffening, as seen in post-tensioned cable mechanisms\cite{boni2021nonlinear}. However, in angle weaves the thread path is contrastingly tortuous, thus subject to significant capstan friction. These frictional effects limit tension propagation, resulting in a reconfigurable structure capable of taking many rigid conformations depending on the arrangements of local geometries~\cite{faber2020dome}. 

In summary, we demonstrated the unique mechanics of beaded materials. We find that the mechanical implications of gravity and friction that often hinder applications of deformable materials can be leveraged, giving way to new opportunities in materials design, robotics, self-assembly, and shape-morphing across length scales. 

\clearpage

\section{Extended data}
\label{link:extdata}

\renewcommand{\figurename}{Extended Data Figure}
\renewcommand{\thetable}{\arabic{table}}
\setcounter{figure}{0}
\setcounter{table}{0}

\newpage
\subsection{Maximum loads of bead shells}
\vspace{-1cm}
\label{link:extdata_shells}
\begin{figure}[!h]
    \centering
    \includegraphics[width=120mm]{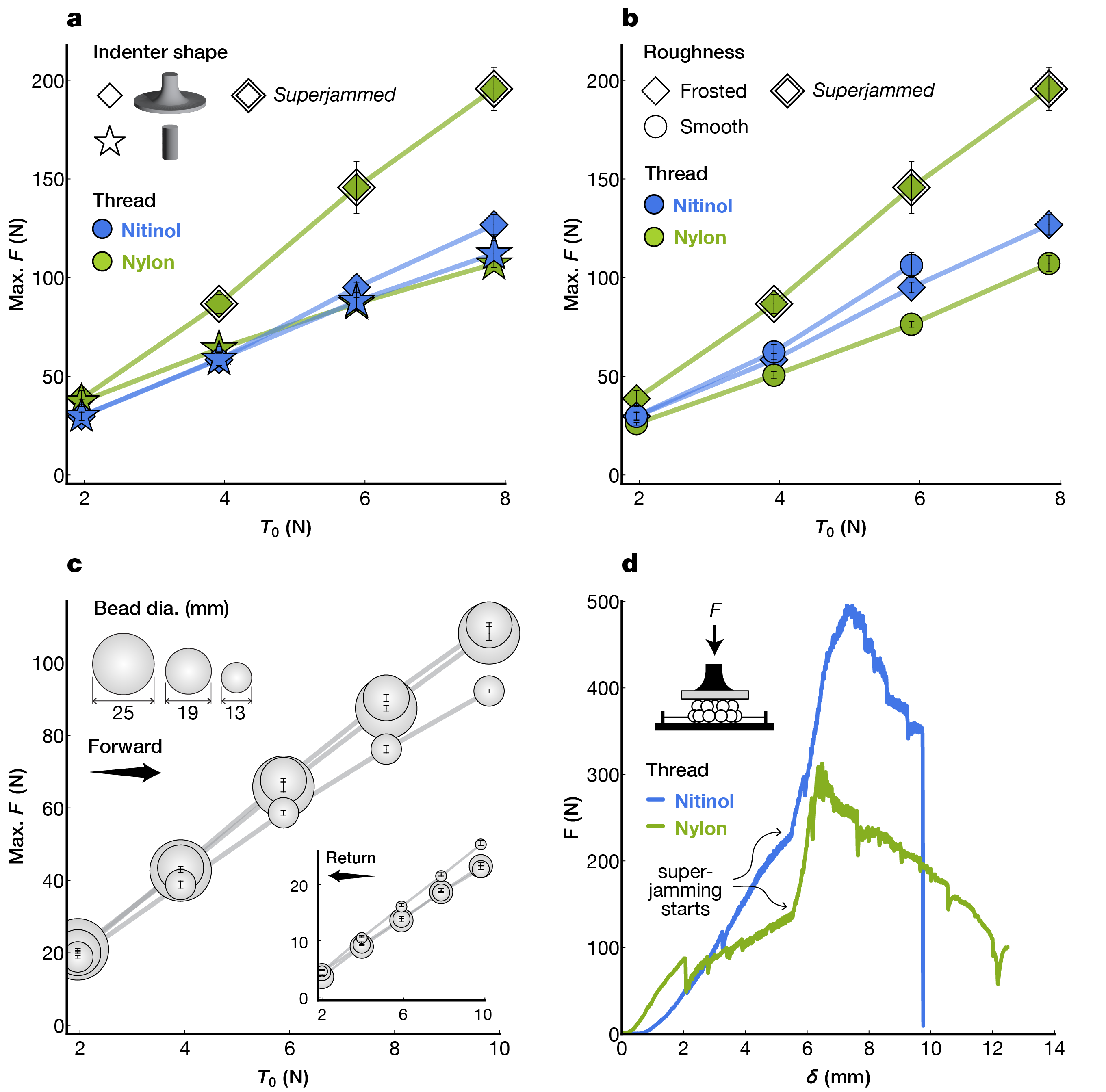}
    \caption{
    \textbf{Maximum loads of bead shells.}
    \textbf{a}, The effect of indenter shape was tested in shells made from frosted, 10-mm acrylic beads threaded with nitinol (blue) and nylon (green), using plate-shaped (diamond) and columnar (star) indenters, where the diameter of the column (17~mm) spans the centers of the upper ring (yellow beads) and does not come into contact with other beads. We report the maximum load across a range of pretensions ($T_0=2.0-7.8$~N), compare to Fig.~\ref{fig:fig2}d. A plate indenter was used for all of the following experiments.
    \textbf{b}, The effect of bead roughness was tested in shells made with 10-mm smooth (circle) and rough i.e. frosted (diamond) acrylic beads threaded with nitinol (blue) and nylon (green). We report the maximum load across a range of pretensions ($T_0=2.0-7.8$~N).
    \textbf{c}, The effect of bead size was tested in shells made with smooth acetal spheres threaded with nitinol and manually drilled to maintain consistent bead hole geometry. We report the maximum load across a range of pretensions ($T_0=2.0-9.8$~N). Inset shows the maximum load during the return.
    \textbf{d}, In lieu of pretension supplied by a sliding weight-and-pulley system, here we manually pretension thread ends and clamp them. We report force vs indenter displacement until failure for shells made with 10-mm frosted acrylic beads threaded with nitinol (blue) and nylon (green). The slopes in the super-jammed regime are 170 and 193~N/mm for nitinol and nylon, respectively.
    }
    \label{fig:fig_extdata_shells}
\end{figure}

\clearpage

\subsection{Motion tracking the 5-bead building block }
\label{link:fig_extdata_forcepath}
\begin{figure}[h!]
    \centering
    \includegraphics[width=120mm]{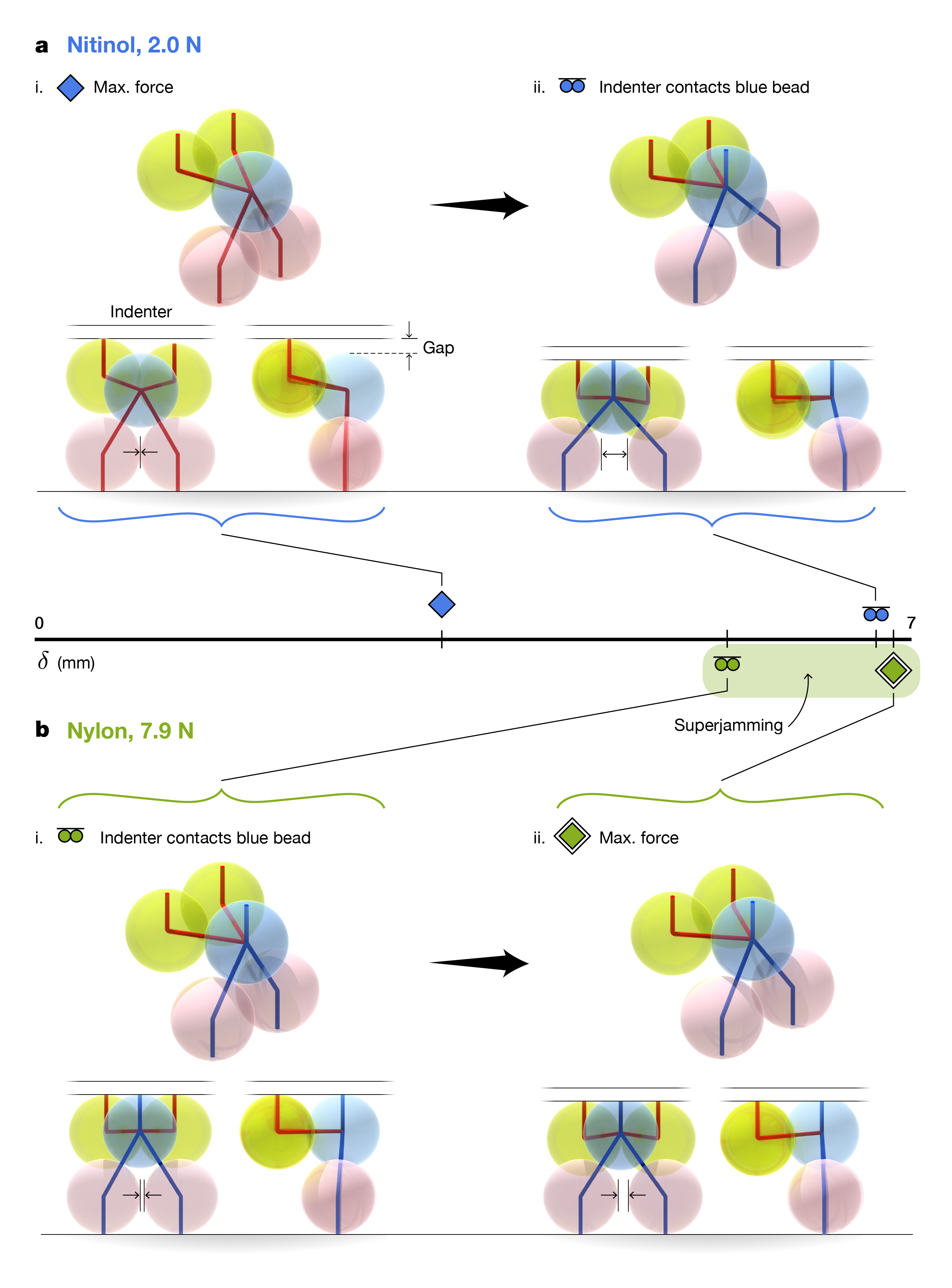}
    \caption{\textbf{3D reconstruction of motion-tracked bead centers.}
    \textbf{ a--b}, Conformations of the 5-bead building block are rendered for shells threaded with nitinol and nylon, tensioned with 2.0 N and 7.9 N, respectively.
    Red and blue lines show the force path, where the indenter contacts the structure at the yellow (upper) and blue (middle) beads, respectively. Superjamming occurs in \textbf{b}. The blue force path is nearly vertical, and the pink beads separate minimally. 
    } 
    \label{fig:fig_extdata_forcepath}
\end{figure}

\clearpage

\subsection{Section cuts through angle weave unit cells}
\label{link:extdata_holes}
\begin{figure}[h!]
    \centering
    \includegraphics[width=120mm]{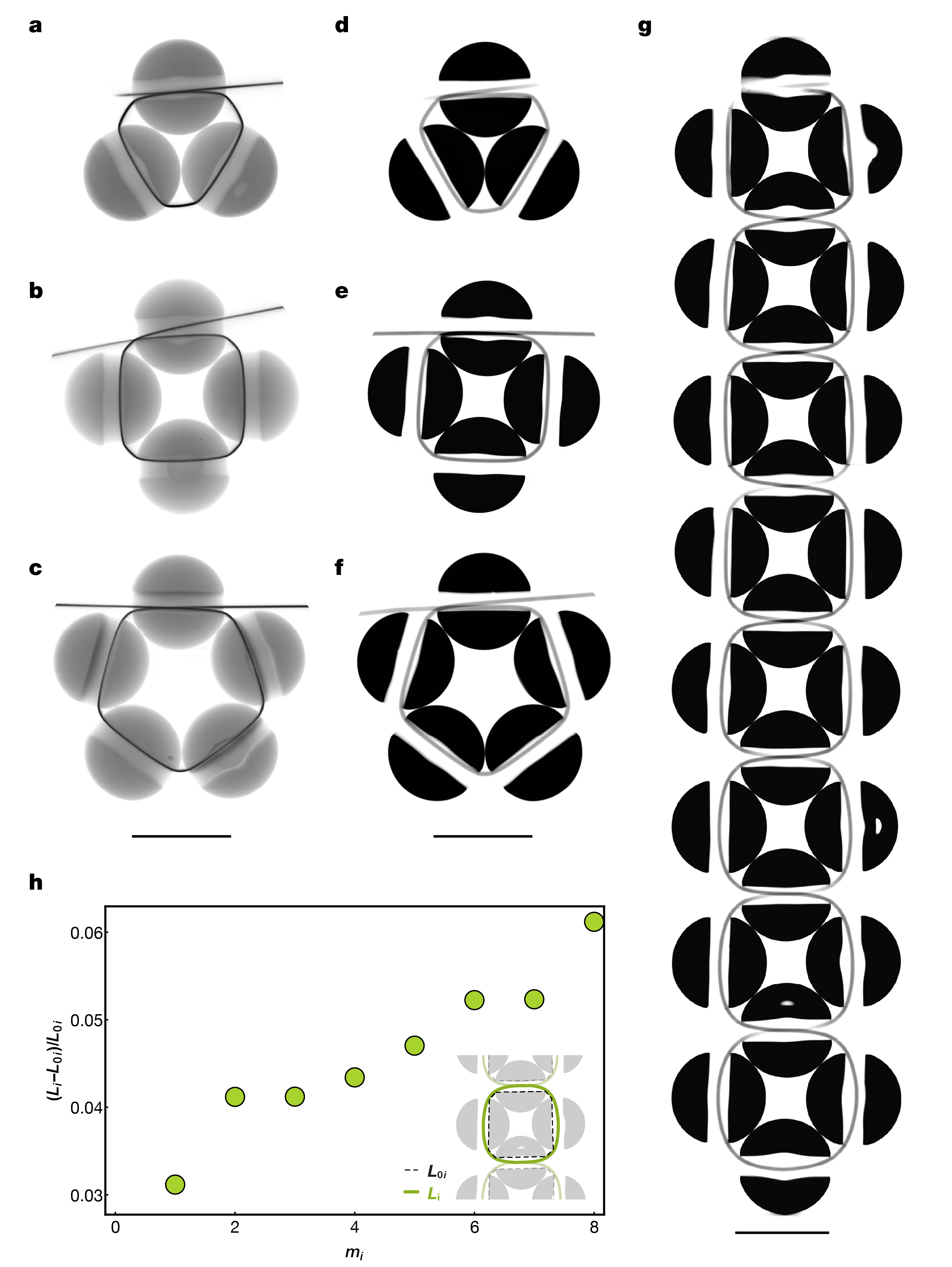}
    \caption{\textbf{Section cuts through angle weave unit cells.}
    \textbf{ a--f},  Section cuts made through $\mu$-CT data at the center of unit cells threaded with nitinol (a, b c) and nylon (d, e, f) for $n=3, 4$ and $5$, respectively. Scale bars, 10~mm.
    \textbf{g}, Continuous chain of $m=8$ unit cells for $n=4$, threaded with nylon. Scale bar, 10~mm.
    \textbf{h},  Nondimensional thread length measured for each ring of panel g.} 
    \label{fig:fig_extdata_rings}
\end{figure}

\clearpage

\subsection{Projected force to dilate cells in a beaded chain}
\label{link:extdata_chain}
\begin{figure}[h!]
    \centering
    \includegraphics[width=120mm]{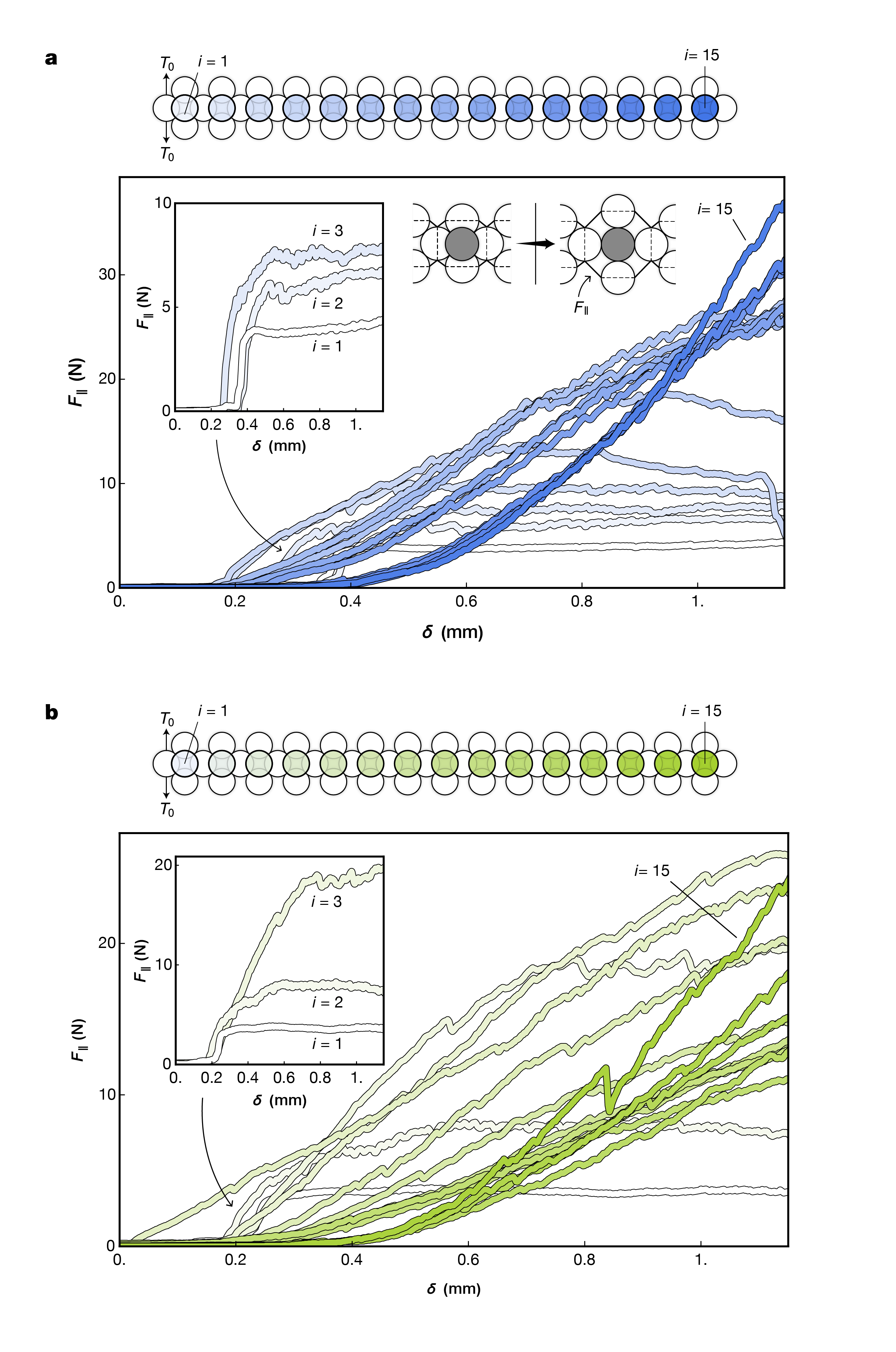}
    \caption{\textbf{Projected force to dilate cells in a beaded chain.}
    \textbf{a--b}, A chain of 15 rings with $n=4$ beads per ring is dilated sequentially with a spherical indenter starting at $i=1$. We report the in-plane projected force $F_{\parallel}$ as illustrated in a inset.  Results are shown for chains with $T_0=2.0$ N, threaded with nitinol and nylon in a and b, respectively. For clarity, inset plots of a and b show the projected force for the three rings closest to the free end. 
    } 
    \label{fig:fig_extdata_chain}
\end{figure}

\clearpage

\subsection{Experimental setup details}
\label{link:extdata_holes}
\begin{figure}[h!]
    \includegraphics[width=\textwidth]{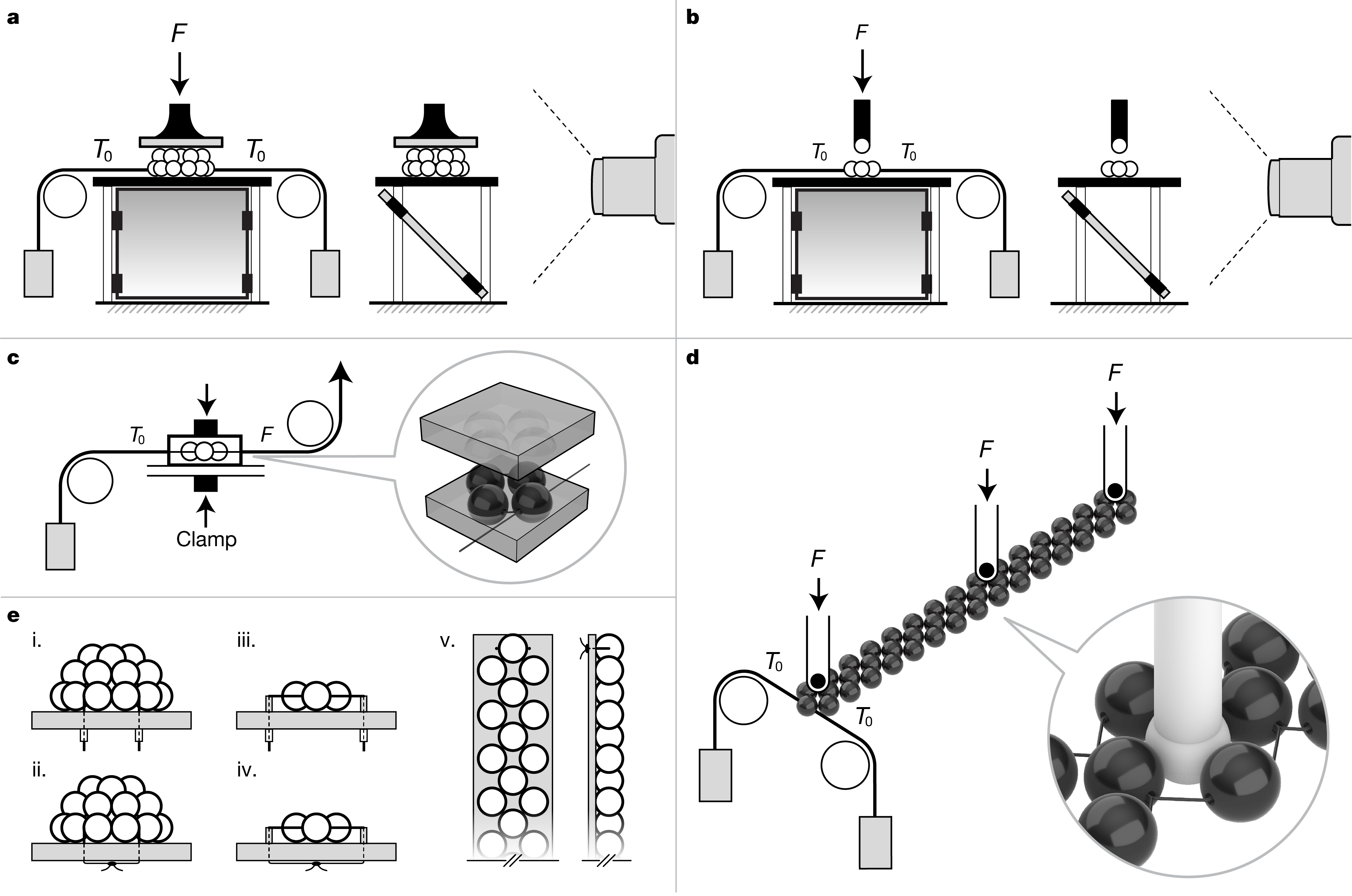}
    \caption{\textbf{Experimental setups details.}
    \textbf{a}, Angled mirror setup for shell compression experiment.
    \textbf{b}, Angled mirror setup for ring dilation experiment.
    \textbf{c}, Measuring capstan friction in a planar ring. Inset shows 3D printed housing into which the ring is inserted (3D rendering); then the assembly is clamped to the experimental stage.
    \textbf{d}, Dilation of a beaded chain where $n=4$ and $m=15$. Inset shows the spherical-tipped indenter (3D rendering).
    \textbf{e}, Tomography setups for the shell (i, ii), rings (iii, iv), and chain where $m=8$ (v), showing clamping techniques. Bead assemblies threaded with nitinol are manually tensioned and crimps are applied to each end. In the case of nylon, the thread is manually tensioned, tied, and glued to prevent slipping. In all cases, the thread is passed through a polymeric platform (3D printed PLA, PETG, or laser-cut acrylic) such that knotting/crimping is performed on the underside. After tensioning, the bead assemblies are glued to the platform.
    }
    \label{fig:fig_extdata_setups}
\end{figure}

\newpage
\clearpage

\subsection{Material information}
\label{link:extdata_materials}
~
\begin{table}[h!]
    \centering
    \begin{tabular}{@{\extracolsep{10pt}}l r r p{4cm} p{4cm}}
        \hline
        \textbf{Material} & \textbf{Dia.} (mm) & \textbf{Hole dia.} (mm) & \textbf{Source} & \textbf{Figure reference}\\
        \hline
         Acrylic, frosted & 5.85 \textpm 0.08 & 1.81 \textpm 0.03 & Amazon.com & \ref{fig:fig4}c\\
         Acrylic, frosted & 9.60 \textpm 0.44  & 2.26 \textpm 0.13 & Amazon.com & \ref{fig:fig1}a; \ref{fig:fig1}b~ii; \ref{fig:fig2}, Ext.~Data~\ref{fig:fig_extdata_shells}a--b, d; SI Video~\ref{v:1} \\
         Acrylic, frosted & 11.54 \textpm 0.06 & 2.52 \textpm 0.06 & Amazon.com & \ref{fig:fig4}a--b\\
         Acrylic, frosted & 17.35 \textpm 0.12 & 2.45 \textpm 0.05 & Amazon.com & \ref{fig:fig1}b~i,iii; SI Video~\ref{v:1}\\
         Acrylic, frosted & 19.29 \textpm 0.14 & 2.45 \textpm 0.05 & Amazon.com & \ref{fig:fig4}d\\
         Acrylic, smooth & 9.80 \textpm 0.05  & 2.16 \textpm 0.05 & Amazon.com & \ref{fig:fig3}; Ext.~Data~\ref{fig:fig_extdata_rings}\\
         Acetal & 25.32 \textpm 0.00 & 1.57 \textpm 0.04 & McMaster-Carr (9614K59) & Ext.~Data~\ref{fig:fig_extdata_shells}c\\
         Acetal & 19.00 \textpm 0.01 & 1.59 \textpm 0.03 & McMaster-Carr (9614K25) & Ext.~Data~\ref{fig:fig_extdata_shells}c\\
         Acetal & 12.66 \textpm 0.00 & 1.59 \textpm 0.03 & McMaster-Carr (9614K27) & Ext.~Data~\ref{fig:fig_extdata_shells}c\\
         Wood & 47.96 \textpm 0.57 & 7.28 \textpm 0.34 & Amazon.com & \ref{fig:fig1}b~iv\\
        \hline
    \end{tabular}
    \caption{Bead materials used in the study}
    \label{tab:beads}
\end{table}

\begin{table}[h!]
    \centering
    \begin{tabular}{@{\extracolsep{10pt}}l r  p{4cm} p{4cm}}
        \hline
        \textbf{Material} & \textbf{Dia.} (mm) & \textbf{Source} & \textbf{Figure reference}\\
        \hline
         Nitinol                      & 0.25 \textpm 0.00  & McMaster-Carr (3617N26), \newline Component Supply (NW-0100) & \ref{fig:fig1}b~ii; \ref{fig:fig2}, \ref{fig:fig3}a--c; Ext.~Data~\ref{fig:fig_extdata_shells}; Ext.~Data~\ref{fig:fig_extdata_rings}a--c; SI Video~\ref{v:1}\\
         Nitinol                      & 0.20 \textpm 0.00  & McMaster-Carr (3617N24)   & \ref{fig:fig3}d\\
         Flexinol+ shape-memory nitinol & 0.20 \textpm 0.00  & Dynalloy & \ref{fig:fig4}c\\
         KastKing monofilament nylon  & 0.50 \textpm 0.01  &  Amazon.com                 & \ref{fig:fig1}a; \ref{fig:fig2}; \ref{fig:fig3}; Ext.~Data~\ref{fig:fig_extdata_shells}a--b, d; Ext.~Data~\ref{fig:fig_extdata_rings}d--h\\
         Coated wire rope             & 0.76 \textpm 0.01 & McMaster-Carr (34235T26)     & \ref{fig:fig1}b~i, iii; SI Video~\ref{v:1}\\
         Coated wire rope             & 0.81 \textpm 0.01 & McMaster-Carr (34235T27)     & \ref{fig:fig1}b~iv\\
         Beadalon "Elasticity" polyurethane    & 0.92 \textpm 0.01 & Amazon.com          & \ref{fig:fig4}a--b\\
         Ritza "Tiger" waxed nylon         & 0.42 \textpm 0.01  & Amazon.com           & \ref{fig:fig4}d\\
        \hline
    \end{tabular}
    \caption{Thread materials used in the study}
    \label{tab:thread}
\end{table}

\newpage
\clearpage

\section{Methods}
\label{link:methods}

\subsection{Materials}
A table listing the materials, dimensions, and sources for bead and thread materials used in this study is provided in Extended Data Tables \ref{tab:beads} and \ref{tab:thread}. 

\subsection{Dodecahedral shell}

\subsubsection{Compression experiments}
\label{link:methods_shell}
In Fig.~\ref{fig:fig1}b iii, the thread ends were manually tensioned and then clamped to a rigid support, as illustrated in Extended Data Fig.~\ref{fig:fig_extdata_shells}d, inset. Thread ends are not visible in the image. Both photographs in Fig.~\ref{fig:fig1}b ii and iii were taken on a  background that was removed. A gray gradient was added at the shell bases. 

Details of the experimental setups used to obtain data reported in Fig.~\ref{fig:fig2}b--d are illustrated in Extended Data Fig.~\ref{fig:fig_extdata_setups}. The setup was fitted with two pulleys and a 45\degree angled mirror positioned beneath a transparent acrylic stage. Masses ranging from 200g--800g were attached to both thread ends, causing the shell to pop out-of-plane due to the closing of the bottom loop with tension. The shell was compressed for three cycles per tension trial using a 3D-printed polylactic acid (PLA) compression plate. Force vs. displacement data was recorded using a universal testing machine (Instron 5940).

\subsubsection{Tracking bead motion}
\label{link:methods_shell_track}
While the shells were compressed, bead positions from the front and bottom were recorded using a digital camera. To obtain the measurements reported in Figs.~\ref{fig:fig2}e-f, motion tracking for a single cycle was performed on bead centers using Adobe After Effects (Adobe Inc.). Force and RGB data were synchronized to the starting motion of the compression plate.
To measure dilation ($\Delta \epsilon_{\text{avg}})$, we fitted circles through the bottom bead centers at each frame, obtaining an approximate dilation radius, $R$. Subtracting the thread length inside of the beads, we found $\Delta \epsilon$ at each video frame such that
\begin{equation}
\Delta \epsilon_{\text{avg}}=\frac{2\pi R-n r^{*}}{m},
\end{equation}
where $r^{*}=9.2$~mm is the average bead diameter taken across bead holes, $n=10$ is the number of beads in the lower ring, and $m=5$ is the number of loops encircling the central loop (yellow beads).

To measure out-of-plane deformation, we tracked x-y-z positions of bead centers in a sample unit cell, taking the mean of the two x-values obtained from front and bottom views. Angles $\theta_{1}$ and $\theta_{2}$ are calculated by taking the mean of the vectors constructed from the blue bead center to each pair of top and bottom beads centers (yellow and pink). For both measurements, we report the altitudinal angle, i.e., the angle formed between these vectors and the x-y plane. 

To find $F_{\parallel}$, we also accounted for in-plane spreading of the pink beads, i.e. $\alpha$ as illustrated in Fig.~\ref{fig:fig2}e. The evolution of $\alpha$ is not shown in Fig.~\ref{fig:fig2}g, but was derived from the vectors used to compute $\theta_2$, taking $1/2$ the angle formed by their projections in the x-y plane. We thus recast $F$, $\theta_2$, and $\alpha$ as functions of $\Delta \epsilon_{\text{avg}}$ and applied the following geometric relation 
\begin{equation}
F_{\parallel}=\frac{F}{m \tan{\theta_2}\sin{\alpha}},
\end{equation}
where $F$ is modulated by angular contacts across $m=5$ building blocks, noting that the effect of $\alpha$ on $F_{\parallel}$ is minimal ($\sin{\alpha}\approx 0.9$--$1.0$).

\subsection{Nitinol and nylon characterization}

\subsubsection{Effective sliding friction, $\mu_\text{eff}$}
\label{link:methods_char_friction}
Rings of $n$ beads were threaded with nitinol and nylon as illustrated in Extended Data Fig.~\ref{fig:fig_extdata_setups}c.
Each assembly was secured between two 3D-printed PLA clamps  designed with hemispherical groves so that the thread could pass unobstructed through the assembly, but the beads remained fixed to the experimental setup when tension was applied. One end of the thread was fed through a frictionless pulley where masses were attached (50--500g in increments of 50g). The other end was fed through another frictionless pulley and passed upwards to a clamp attached to a universal testing machine. The thread at the sensor was moved upwards at 30 mm/s, and we recorded the tension vs. displacement at this end.
We repeated the experiment across bead assemblies of different sizes: $n=3$, $4$, $5$, and $6$.

\subsubsection{Young's modulus, $E$}
\label{link:methods_char_thread}
We measured the force $F$ to stretch thread samples of different lengths a distance $\delta$ axially when secured between two clamps. Trials were conducted with clamps spaced 10, 25, 50, and 100~mm apart, with three trials performed at each distance. To find the Young's modulus $E$, we extracted a slope $s$ fitted to the elastic regime of each $F$ vs. $\delta$ displacement curve. We then calculated $E$ such that $E = s L_0 / A$, where $A$ is the unstretched thread cross-section and $L_0$ incorporates a fitted, additional length assumed to be stretching inside of each clamp (13~mm and 9~mm for nitinol and nylon, respectively, per clamp). Reporting the mean $E$ across samples, we obtained \( 38.2 \pm 3.2 \) and \(1.80 \pm 0.13\) GPa for nitinol and nylon, respectively. We note that both materials remained linearly elastic until $\delta/L_0 \approx 0.015$.

\subsection{Ring dilation}

\subsubsection{Experiments}
\label{link:methods_dilate_exp}

To obtain the dilation data shown in Fig.~\ref{fig:fig3}c, we assembled beaded rings identical to those tested in Figs.~\ref{fig:fig3}a--b and measured the force to dilate the rings using a spherically-tipped indenter (see Extended Data Fig.~\ref{fig:fig_extdata_setups}b), cycling three times for a range of pretensions ($T_0=2.0$--$9.8$ N).
In Fig.~\ref{fig:fig3}d, we constructed a chain of $m=15$ unit cells where $n=4$ via a "two-handed right-angle-weave" (RAW) design. Bead rings are linked by a single common bead and two thread ends are worked in an alternating fashion such that the thread snakes through the weave on alternating sides of each ring. In this manner, the thread accumulates a capstan wrapping angle of $\phi = \pi$ per ring. In Extended Data Fig.~\ref{fig:fig_extdata_rings}g we show a $\mu$-CT section cut through a RAW chain where $m=8$. To dilate the chain, a similar approach was used as described previously, using a spherical-tipped indenter. The chain was dilated sequentially starting at the free end ($m=1$), moving towards $m=15$ (see Extended Data Fig.~\ref{fig:fig_extdata_setups}d), performing three dilation cycles per ring over a range of $T_0$ as reported in Fig.~\ref{fig:fig3}d.

\subsubsection{Model}
\label{link:methods_dilate_model}

We dilate a ring of $n$ beads with radius $a$ with a spherical indenter of the same radius. The geometric relationship between indenter displacement $\delta$ and the altitudinal angle $\theta$ is described as
\begin{equation}
    \delta=2a(1-\sin{\theta}),
\end{equation}
where $\theta$ is defined as the angle between the line connecting the indenting sphere center and the center of a bead in the ring and the x-y plane.
The transfer of vertical force to the bead ring occurs through bead contacts such that
\begin{equation}
    \frac{F}{n}= F_{n} \sin{\theta},
\end{equation}
where $F_{n}$ represents the contact force between a single bead and the indenter. Projected in the x-y plane, the component dilation force $F_{n}\cos{\theta}$ is resisted by threads that emerge from bead holes at an angle $\beta$. We proceed with the assumption that tension in the thread is constant and, as a result, that the ring deforms with radial symmetry. In static equilibrium, we can write the balance of in-plane forces along radial lines of symmetry as
\begin{equation}
    2T\cos{\beta}= F_{n}\cos{\theta},
\end{equation}
where $\beta$ is constant due to assumed symmetry and can be derived from the regular polygon formed by discrete bead edges such that $\beta=\frac{\pi}{2}-\frac{\pi}{n}$. We solve this system of three equations and plot the results in Fig. \ref{fig:fig3}c.

To overlay experimental data with the model, we define our reference for $\delta = 0$ by assuming the point of maximum force occurs when the indenter makes contact with a perfectly taut bead ring. In-plane, we can describe the radius $R$ of a circle that intersects the centers of a taut loop of $n$ beads as 
\begin{align}
    R = \frac{a}{\sin(\pi/n)}.
\end{align}
Out-of-plane, we can thus write the contact height $h=2a-\delta$ of the indenting sphere as 
\begin{align}
h = a \left( 4 - \frac{1}{\sin^2\left(\pi/n\right)} \right) ^{1/2}. 
\end{align}

\subsection{Larger bead network experiments}

\subsubsection{Column beaded with elastomeric thread}
\label{link:methods_larger_col}
We used a “prismatic right-angle weave" (PRAW) technique to achieve the columnar geometry, consisting of angle weave loops of $n=4$ oriented vertically. These loops are constructed around an initial ring of a given size --- here, also $n=4$ --- making this a "PRAW-4" design. We constructed 11 layers using 12-mm acrylic beads with elastomeric thread (see Extended Data Tables \ref{tab:beads} and  \ref{tab:thread}) to create a square prism, 92~mm long at rest length. We manually tensioned the thread at the free end and tied a knot to maintain tension. 
The column was secured to a universal testing machine using a custom clamping apparatus (3D printed PETG, laser-cut acrylic, hardware) that grips the end beads but not the thread (see Fig.~\ref{fig:fig4}a, inset). We recorded the force to compress and extend the column, cycling three times over a total displacement of 35~mm extension and 23~mm compression at 30~mm/min.

\subsubsection{Beam beaded with SMA wire}
\label{link:methods_larger_beam}
We beaded a beam using a PRAW-3 design (see above) consisting of 31 layers of 6-mm acrylic beads and shape-memory nitinol (see Extended Data Tables \ref{tab:beads} and  \ref{tab:thread}) to create a triangular prism, 125~mm long at rest length. We manually tensioned the thread at the free end and applied metal crimps to maintain tension. 
Alligator clips connected to a benchtop power supply were connected to exposed nitinol at both beam ends.  The “on” voltage setting (6V) was selected by increasing the voltage at the power supply until the current draw reached the maximum value recommended by the manufacturer per the wire diameter (660~mA)\cite{dynalloy}. 
We then conducted 3-point bending tests in both “off” and “on” states, cycling a total of 9 and 12 times, respectively, at 30~mm/min. Linear models were fitted to the resulting force vs. displacement plots.

\subsubsection{Catenary shell}
\label{link:methods_larger_shell}
A catenary shell geometry was generated in 3D modeling software (\textit{Rhino/Grasshopper}, Robert McNeel \& Associates). We then constructed a triangular mesh with optimized edge lengths by packing spheres of equal radii on this surface \cite{piker2017circle} and connecting each point to six nearest neighbors. 
We placed beads at mesh face centers to generate a beadable pattern, which was then assembled by hand using 20-mm acrylic beads and a waxed, filamentous nylon thread (see Extended Data Tables \ref{tab:beads} and \ref{tab:thread}). 
In the photographs of Fig~\ref{fig:fig4}d, the shell was manipulated into different conformations and set to rest with gravity. The shell was photographed on a black background that was removed.

\subsection{Tomography}
\label{link:methods_tomography}
$\mu$-CT imaging in Figs~\ref{fig:fig1}b ii, \ref{fig:fig4}b, and \ref{fig:fig_extdata_rings}a--g was conducted using a Zeiss Xradia Versa 520 3D X-ray microscope. Tension was applied to threads by manually pulling the ends and tying a knot. Tomography data was rendered using Dragonfly 3D visualization software (Comet Technologies). Fig.~\ref{fig:fig4}b (beaded column) and Extended Data Figs. \ref{fig:fig_extdata_rings}a--c (nitinol samples) show 3D rendered orthographic views of a cropped region of interest. Extended Data Figs. \ref{fig:fig_extdata_rings}d--g (nylon samples) use a "mean slab" section cut across a thickness of approximately 1~mm.

\newpage
\section{Acknowledgements}
This work was supported by NSF Future Manufacturing Grant CMMI 2037097. 
We would like to thank ballet dancer Gigi Schadrack for standing on the experiments in Fig.~\ref{fig:fig1}b iii--iv. TJJ would like to thank Tracey Jones for introducing him to the art of bead-weaving.

\renewcommand\thesubsection{\arabic{subsection}}
\setcounter{section}{0}
\setcounter{subsection}{0}
\section*{Supplementary Information}

\subsection{Supplementary Video 1}
\label{v:1}
3D geometry of a beaded shell. Applying tension to thread ends sends the half-dodecahedral shell out-of-plane, as pictured in Fig.~\ref{fig:fig1}. This is followed by a 360\degree rotation of reconstructed micro-computed tomography ($\mu$-CT) data of the shell studied in Fig.~\ref{fig:fig2}. The shells are identical in design but use different thread materials and bead sizes (see Tables~\ref{tab:beads}--\ref{tab:thread}). 

\subsection{Supplementary Video 2}
\label{v:2}
Videos of the shell from Fig.~\ref{fig:fig2} as it is compressed, shown with motion-tracked points and corresponding force data. 

\end{document}